\documentclass[10pt,twocolumn,superscriptaddress,aps,prl,balancelastpage]{revtex4-1}
\usepackage{amsfonts,amsmath,amssymb,graphicx,epstopdf,verbatim,dsfont,xcolor}
\usepackage[english]{babel}
\usepackage{subfigure}
\usepackage{epstopdf}
\usepackage{hyperref}
\hypersetup{
  colorlinks = true,
  citecolor = blue
}

\usepackage{physics}
\usepackage[all]{hypcap}

\usepackage{cleveref}

\newcommand{\be}{\begin{equation}}
\newcommand{\ee}{\end{equation}}
\newcommand{\<}{\langle}
\renewcommand{\>}{\rangle}

\newcommand{\beginsupplement}{%
        \setcounter{table}{0}
        \renewcommand{\thetable}{S\arabic{table}}%
        \setcounter{figure}{0}
        \renewcommand{\thefigure}{S\arabic{figure}}%
        \setcounter{equation}{0}
        \renewcommand{\theequation}{S\arabic{equation}}
     }

 \graphicspath{{figs/}}
 
\begin{document}

\title{Monitoring-induced Entanglement Entropy and Sampling Complexity}
\author{Mathias Van Regemortel}
\email{mvanrege@umd.edu}
\affiliation{Joint Quantum Institute and Joint Center for Quantum Information and Computer Science,
NIST/University of Maryland, College Park, Maryland 20742, USA}

\author{Oles Shtanko} 
\affiliation{IBM Quantum, IBM Research -- Almaden, San Jose CA, 95120, USA}

\author{Luis Pedro Garc\'ia-Pintos}
\affiliation{Joint Quantum Institute and Joint Center for Quantum Information and Computer Science,
NIST/University of Maryland, College Park, Maryland 20742, USA}

\author{Abhinav Deshpande}
\affiliation{Institute for Quantum Information and Matter, California Institute of Technology, Pasadena, California 91125, USA}

\author{Hossein Dehghani}
\affiliation{Joint Quantum Institute and Joint Center for Quantum Information and Computer Science,
NIST/University of Maryland, College Park, Maryland 20742, USA}

\author{Alexey V. Gorshkov} 
\affiliation{Joint Quantum Institute and Joint Center for Quantum Information and Computer Science,
NIST/University of Maryland, College Park, Maryland 20742, USA}

\author{Mohammad Hafezi }
\affiliation{Joint Quantum Institute and Joint Center for Quantum Information and Computer Science,
NIST/University of Maryland, College Park, Maryland 20742, USA}

\date{today}

\begin{abstract}
The dynamics of open quantum systems is generally described by a master equation, which describes the loss of information into the environment. By using a simple model of uncoupled emitters, we illustrate how the recovery of this information depends on the monitoring scheme applied to register the decay clicks. The dissipative dynamics, in this case, is described by pure-state stochastic trajectories and we examine different unravelings of the same master equation. More precisely, we demonstrate how registering the sequence of clicks from spontaneously emitted photons through a linear optical interferometer induces entanglement in the trajectory states. Since this model consists of an array of single-photon emitters, we show a direct equivalence with Fock-state boson sampling and link the hardness of sampling the outcomes of the quantum jumps with the scaling of trajectory entanglement.

\end{abstract}

\maketitle

The coupling of a quantum system to an environment generally leads to decoherence and, under certain conditions, can be modeled by a Markovian master equation that could generically result in a mixed (non-pure) density matrix \cite{gardiner2004quantum}. An alternative but equivalent approach describes the ``unraveling'' of the same density matrix in terms of pure-state stochastic wave-function trajectories \cite{dalibard1992wave,dum1992monte,carmichael1993quantum,weimer2021simulation}. Interestingly, for a given master equation, the unraveling in terms of stochastic trajectories is not unique. For example, note that a Lindblad master equation,
\begin{equation}
\label{eq:ME}
    \partial_t \rho = \gamma \sum_j \Big( c_j \rho c^\dagger_j -\frac 12 \{ c^\dagger_j c_j, \rho \} \Big),
\end{equation}
is invariant under any transformation $c_i \rightarrow \sum_j U_{ij} c_j$, where $U$ is a unitary matrix and $\gamma$ is the decoherence rate. Here, $c_j$ are the jump operators that describe dissipative coupling to the environment (see \cite{supp}). In particular, this implies that any observable $\langle O \rangle = \text{Tr} (\rho O)$ preserves its expectation value, independent of the choice of $U$. In the unraveling picture, on the other hand, the unitary $U$ is of direct importance for the stochastic quantum states, as can be understood by evaluating the effect of a quantum jump $c_i\ket{\psi}$. Nevertheless, averaging expectation values over different trajectory states will converge back to the $U$-independent result from the master equation, $\mathbb E_\psi \<\psi|O|\psi\> = \Tr (\rho O)$, where $\mathbb E_\psi$ is the expectation over all individual trajectories $|\psi\>$. This is in contrast with the case of nonlinear quantities, such as bipartite entanglement entropy, which may show an unraveling dependence.

Physically, the specific choice of unraveling of a master equation is determined by the physical observable that is monitored in a dissipative process~\cite{cirac1994quantum,buhner2000resonance,nha2004entanglement,kok2007linear,gambetta2008quantum,wiseman2012dynamical}, e.g.~detecting the decay of a two-level system by observing the emitted single photon. Remarkably, such stochastic quantum trajectories were observed in several pioneering experiments in trapped-ion systems \cite{bergquist1986observation,nagourney1986shelved,sauter1986observation,leibfried2003quantum} and circuit quantum electrodynamics (circuit-QED) \cite{murch2013observing}. Moreover, it has been shown that monitoring such trajectories can be used to manipulate stochastic quantum systems ~\cite{weber2014mapping,sun2014tracking,hacohen2018incoherent,ficheux2018dynamics,flurin2020using}, with potential applications in quantum error correction~\cite{akerman2012reversal,minev2019catch}.

Furthermore, from a theoretical perspective, monitoring may have a profound impact on the stochastic trajectory states when it competes with coherent processes. Specifically, it was shown that a scaling transition for averaged trajectory entanglement entropy can occur~\cite{cao2019entanglement,alberton2021entanglement,fuji2020measurement,van2021entanglement}.
In these works, dissipation was studied in the context of a measurement-induced phase transition~\cite{nahum2017quantum,skinner2019measurement},
and the master equation associated with the dissipative dynamics was 
changing across the phase transition. 
This implies that the effect of the monitoring protocol itself and the corresponding choice of unraveling remain largely unexplored for the scaling of entanglement entropy in the stochastic trajectory states.

\begin{figure}
    \centering
    \includegraphics[width=\columnwidth]{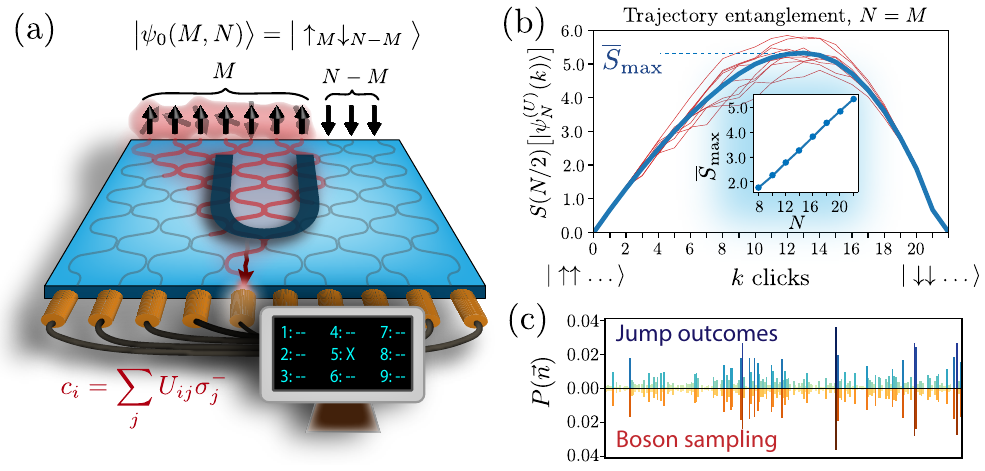}
    \caption{ (a) A schematic illustration of the setup, consisting of a chain of $N$ two-level emitters, $M$ of which are initially in the $\ket{\uparrow}$ state, with the remaining $N-M$ in the $\ket{\downarrow}$ state. The quantum jumps from the spontaneous emissions in the chain are monitored through the output ports of a linear optical network 
    represented by an $N\times N$ unitary $U$, 
    giving new jump operators $c_i$. (b) The case $N=M=22$ and $U$ sampled from the $N\times N$ Haar measure: half-chain entropy for some stochastic trajectories (red) and the averaged value (blue). The inset shows the volume-law scaling of the maximal averaged entanglement entropy $\overline{S}_\text{max}$. (c) After registering $M$ clicks, the jump outcome probabilities are given by Fock-state boson sampling from Eq.\,(\ref{eq:boson_sampling}). A comparison for $N=7$, $M=4$, giving $210$ possible outcomes, and a Haar-random $U$, sampled with $10\,000$ quantum trajectories from the associated unraveling.}
    \label{fig:Fig1}
\end{figure}

\begin{figure}
    \centering
    \includegraphics[width=\columnwidth]{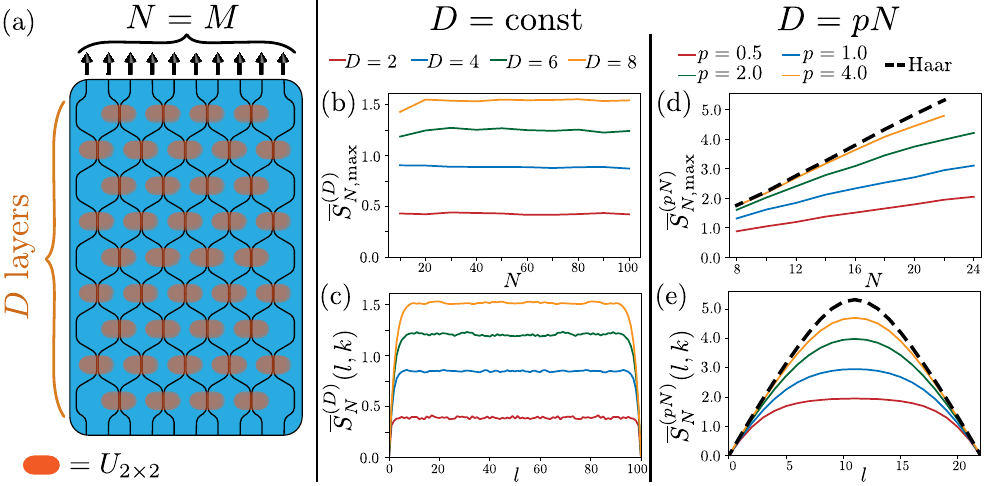}
    \caption{ The entanglement generated in the chain of emitters is studied by monitoring the decays through a LON. (a) Schematic of the setup, where $N$ emitters are excited and monitored through a $D$-layered LON consisting of staggered layers of $2\times 2$ Haar random unitaries from Eq.~\eqref{eq:U2by2}. (b)-(c) A network of constant depth $D$ shows area-law scaling: (b) Increasing $N$, $\overline{S}^{(D)}_\text{max}$ remains stable, and (c) Entanglement profiles $\overline{S}^{(D)}_N(l,k_\text{max}(D))$ for $N=100$, selected after $k_\text{max}$, when the maximal entropy is reached, saturate in the bulk (we find that $k_\text{max}$ is independent of $l$). (d)-(e) Taking $D$ to scale with system size as $D=pN$ gives a volume law for entropy, converging to the result of an $N\times N$ Haar-random unitary for large $p$. (d) Scaling of $\overline{S}^{(D)}_\text{max}$ with system size show linear growth, and (e) the profiles $\overline{S}^{(D)}_N(l,k_\text{max}(p))$ for $N=22$ show a strong dependence on subsystem size $l$.}
    \label{fig:Fig2}
\end{figure}

In this Letter, we consider different monitoring schemes that correspond to different unravelings of the same master equation and analyze the associated impact on stochastic quantum dynamics. We consider an array of uncoupled single-photon emitters whose decay can be monitored by detected photons. A linear optical network (LON) is positioned between the emitters and the detectors, as shown in Fig.~\ref{fig:Fig1}(a), so that the new jump operators correspond to a LON-determined linear combination of the decay jump operators. As the sequence of jump clicks is recorded, a buildup and decay of entanglement entropy is generated in the state of the emitters ---see Fig.~\ref{fig:Fig1}(b). When the LON unitary is Haar random (see e.g. \cite{collins2006integration}), the averaged entanglement entropy reaches a maximum over time that has  volume-law scaling, as shown in the inset of Fig.~\ref{fig:Fig1}(b). Moreover, since a series of single-photon emissions is recorded, we analytically verify a direct equivalence between sampling the outcomes of the decay jumps and the Fock-state boson sampling problem \cite{aaronson2011computational}, as we also numerically demonstrate in Fig.~\ref{fig:Fig1}(c). Finally, we illustrate in Fig.~\ref{fig:Fig2} that the depth of the LON determines the scaling of maximal trajectory entanglement entropy over time, ranging from area law for constant depth to volume law when the depth is proportional to the number of emitters. Given the connection of our system to Fock-state boson sampling, we relate the scaling of maximal trajectory entanglement entropy to the hardness of classically sampling the jump-outcome probabilities: polynomial vs. superpolynomial time, respectively \cite{temme2012efficient,lundow2022efficient}. Utilizing the setup described above, we therefore establish clear connections between the invariance properties of the master equation, the scaling of the associated trajectory entanglement entropy, and the sampling complexity of jump outcomes.

\textit{The model.}---Our setup consists of a chain of $N$ two-level systems that emit photons via de-excitation and are monitored through the output arms of a LON, represented by an $N\times N$ unitary $U$. We start from a state with $M$ two-level systems in the excited state $\ket{\uparrow}$ and $N-M$ in the ground state $\ket{\downarrow}$, i.e. \mbox{$|\psi_0(M,N)\rangle \equiv |\uparrow_1\dots \uparrow_M \downarrow_{M+1}\dots \downarrow_{M-N}\rangle$}, and assume a uniform rate $\gamma$ for the excited emitters to spontaneously emit a photon and relax to the ground state, as depicted in Fig.~\ref{fig:Fig1}(a). 

It is assumed that $\tau_d \ll 1/(M\gamma)$, with $\tau_d$ comprising the time for a photon to traverse the LON and the detector dead time. A jump click recorded in output arm $i$ of the LON $U$ now corresponds to applying the jump operator
\begin{equation}
\label{eq:jump_U}
    c_i \equiv \sum_{j=1}^{N} U_{ij} \sigma^-_j,
\end{equation}
with $\sigma^-_j=(\sigma^x_j-i\sigma^y_j)/2$ the decay operator of emitter $j$ and $\sigma^{x,y,z}_j$ the Pauli $(x,y,z)$-operator acting on site $j$. 

As was emphasized earlier and shown in more detail in Ref.~\cite{supp}, the Lindblad master equation, given by $\partial_t \rho =\gamma \sum_i \big( \sigma^-_i \rho \sigma^+_i - \frac{1}{2}\{\sigma^+_i\sigma^-_i, \rho \} \big)$, is invariant under unitary mixing of the jump operators \eqref{eq:jump_U}. On the level of the master equation, the dynamics of the (uncoupled) emitters is a simple classically mixed state, for which the single-emitter density matrix entries evolve for each emitter independently as $\rho_{\uparrow\uparrow} = 1-\rho_{\downarrow\downarrow} = e^{-\gamma t}$, $\rho_{\downarrow\uparrow}=\rho_{\uparrow\downarrow}=0$ with $\rho_{ij} = |i\rangle \langle j|$.

\textit{Stochastic quantum trajectories.}---A crucial element in this work is the explicit monitoring and recording of the jumps $c_i$ \eqref{eq:jump_U}. The stochastic dynamics resulting from registering the photon clicks in the output arms of $U$ can be simulated with pure-state trajectories \cite{dalibard1992wave,dum1992monte, carmichael1993quantum}. Given a state $\ket{\psi(t)}$, we evaluate the probability for jump $c_i$ to occur in a short time interval $\Delta t$ as $p_i(t) = \gamma \Delta t \bra{\psi(t)}c^\dagger_i c_i\ket{\psi(t)}$. The probability $p_\text{jump}(t) = \sum_i p_i(t)$
determines whether a jump happens at time $t$ or not. If a jump happens, then $c_i$ is selected with probability $\propto p_i(t)$, and we evaluate $\ket{\psi(t+ \Delta t)} = c_i\ket{\psi(t)}$. If there is no jump, the system evolves for time $\Delta t$ under the effective non-Hermitian Hamiltonian $H_\text{eff}=-\frac{i\gamma}{2} \sum_j c_j^\dagger c_j$. In both scenarios, the state is renormalized after each time step. In the limit $\Delta t\rightarrow 0$, averaging $\langle O \rangle$ over sampled trajectory states is equivalent to computing  $\langle O \rangle$ via the master equation~\eqref{eq:ME}.

Note that $H_\text{eff}$ only depends on the number of excited emitters $N_\text{exc} = \sum_i \sigma^+_i \sigma^-_i = \sum_j c_j^\dagger c_j$, and that $\ket{\psi(t)}$ is an eigenstate of $N_\text{exc}$ between jumps if we start from $|\psi_0(N,M)\rangle$. This means that, after renormalization, the evolution between jumps does not change the stochastic state $\ket{\psi(t)}$. 

For the rest of the work, we will therefore discard the explicit time dimension and express the evolution in terms of the jump sequence $(m_1,\dots,m_M)$, with $m_k$ representing the $k$th click in output arm $1\leq m_k \leq N$ and $1\leq k\leq M$. This sequence can be obtained reliably when $\tau_d\ll1/(M\gamma)$, since the photon clicks are now registered with an accuracy significantly higher than the duration of emission (the temporal extent of the photonic wavepacket).

\textit{Connection to remote entanglement of two emitters.}---To intuitively explain the idea and illustrate the underlying correspondence with bosonic statistics, we start with the simple case of two excited emitters and a $2\times2$ LON ($N=M=2$) parametrized as
\begin{equation}
\label{eq:U2by2}
    U = \left(
    \begin{array}{cc}
       a & b \\
        -e^{i\phi}b^\ast & e^{i\phi}a^\ast
    \end{array} \right),
\end{equation}
with $|a|^2+|b|^2=1$, quantifying the mixing between the modes, and $\phi$ the relative phase shift. Setting $a=b=1/\sqrt{2}$ and $\phi=\pi$, corresponding to a $50:50$ beam splitter, gives two new jumps $c_s = \frac{1}{\sqrt{2}}\left( \sigma^-_1 + \sigma^-_2\right)$ and $c_a = \frac{1}{\sqrt{2}}\left( \sigma^-_1 - \sigma^-_2\right)$, the symmetric and antisymmetric jump, respectively.
In case a symmetric click is observed, the symmetric jump $c_s$ is applied to the initial state $\ket{\uparrow\uparrow}$, giving the symmetric Bell state $\ket{\psi_s} = \frac{1}{\sqrt{2}}(\ket{\uparrow \downarrow } +\ket{\downarrow \uparrow })$. This state can only decay another time with the same symmetric jump $c_s$, as seen immediately by evaluating the probabilities $P_i \propto \langle \psi_s| c_i^\dagger c_i|\psi_s \rangle$, with $i=(a,s)$. The same story holds for the antisymmetric jump $c_a$, and, therefore, upon monitoring the output arms of the beam splitter, either the jump sequence $(m_s,m_s)$ or $(m_a,m_a)$ is detected, each with probability $\frac{1}{2}$, and never the sequence $(m_s,m_a)$ or $(m_a,m_s)$. This is equivalent to the celebrated Hong-Ou-Mandel effect for two indistinguishable photons, incident on the two input arms of a $50:50$ beam splitter \cite{hong1987measurement}. In our case, however, the indistinguishable photonic wavepackets are detected after a time much shorter than the duration of emission. As a result, an intermediate maximally entangled (anti)symmetric Bell state between the two emitters is established to convey the interference between the emitted photons. A similar procedure was considered to generate entanglement between cold atoms in a lattice configuration \cite{elliott2015multipartite} and experimentally implemented to entangle two distant trapped ions ~\cite{moehring2007entanglement}. The effect can also be viewed as superradiant emission~\cite{wiegner2015simulating}.

\textit{Correspondence with boson sampling}---We now generalize the system to $N$ emitters, of which $M$ are excited, and an $N\times N$ unitary $U$, representing the LON with monitored output arms---see Fig.~\ref{fig:Fig1}(a). After having registered all $M$ clicks, an observer knows that all emitters have reached the ground state $\ket{\psi} = \ket{\downarrow\downarrow \dots}$. The probability of detecting the $M$ clicks in the Markovian sequence $\vec{m}\equiv (m_1,m_2,\dots,m_M)$ can be evaluated as (see \cite{supp})
\begin{eqnarray}
\nonumber
    P(\vec{m}) &=& 
    \frac{1}{M!} \big\langle \psi_0(M,N) \big| c^\dagger_{m_1}\dots c^\dagger_{m_M}
  \,c_{m_M}\dots c_{m_1}\big|\psi_0(M,N) \big\rangle \\\nonumber
  &=& \frac{1}{M!} \sum_{\vec{k},\vec{l}} U^\ast_{m_1,k_1}\dots U^\ast_{m_M,k_M}\, U_{m_M,l_M}\dots U_{m_1,l_1}\\\nonumber
  && \times \big\langle \psi_0(M,N) \ket{ \sigma^+_{k_1}\dots \sigma^+_{k_M} \,  \sigma^-_{l_M}\dots \sigma^-_{l_1} \big| \psi_0(M,N)} \\
  &=& \frac{|\text{Per}\big( U_T \big)|^2}{M!}.
      \label{eq:markovian_clicks}
  \label{eq:seq}
\end{eqnarray}
Here, $\text{Per}\big(A\big) = \sum_{\sigma \in S_M} \prod_{i=1}^M A_{i,\sigma(i)}$ is the permanent of an $M \times M$ matrix $A$, with $S_M$ the symmetric group, i.e. the summation is performed over the $M!$ possible permutations of the numbers $1,\dots,M$. $U_T$ is the $M\times M$ matrix constructed from $U$ by taking the first $M$ columns and repeating the $i$th row $n_i$ times, where $n_i$ is the number of times detector $i$ appears in the sequence $\vec m$. $|\text{Per}\big( U_T \big)|^2$ arises from gathering all terms that give unit (nonzero) expectation value in the second line of Eq.~\eqref{eq:markovian_clicks}. Expression \eqref{eq:markovian_clicks} can also be obtained with multi-boson correlation sampling, i.e.~by evaluating the $M$th-order temporal correlation function of the photonic quantum state at the output ports of the LON \cite{tamma2016multi}. 

We see that $P(\vec m)$ is the same for all $\vec m$ that give rise to a given $\vec n=(n_1,\dots, n_N)$. Therefore, the probability of registering clicks $\vec n$ with $\sum_i n_i = M$ is obtained simply by multiplying the expression \eqref{eq:markovian_clicks} by the number of sequences $\vec m$ that give rise to this $\vec n$, so that
\begin{equation}
  P(\vec{n}) = \frac{|\text{Per}\big( U_T \big)|^2}{\prod_i n_i!}.
  \label{eq:boson_sampling}
\end{equation}

 The jump outcome probabilities $P(\vec{n})$ in Eq.~\eqref{eq:boson_sampling} are exactly the ones found for Fock-state (conventional) boson sampling when $M$ indistinguishable photons are sampled after passing through an $N \times N$ interferometer~\cite{aaronson2011computational,broome2013photonic}, as verified in Fig.~\ref{fig:Fig1}(c). When $U$ is drawn from the Haar measure and $N=O(M^2)$, it has been proven that sampling from the output distribution is classically hard (takes superpolynomial time) unless the Polynomial Hierarchy collapses to the third level. This follows from the $\#$P-hardness of classically computing the output probabilities in Eq.~\eqref{eq:boson_sampling}.

Experimentally, Fock-state boson sampling has been implemented for small numbers of photons, well within the classically simulable regime \cite{tillmann2013experimental,spring2013boson,spagnolo2014experimental}. Gaussian boson sampling \cite{hamilton2017gaussian}, using squeezed states instead of single photons as input, can be scaled up further, leading to one of the first claims of experimental quantum advantage \cite{zhong2020quantum}.  
Interestingly, by engineering long-range interactions, Fock-state boson sampling was also proven to be equivalent to sampling spin measurement outcomes after a short Hamiltonian time evolution~\cite{olivares2016quantum,peropadre2017equivalence}.

\textit{Trajectory entanglement entropy.}---Our primary interest lies in evaluating nonlinear properties of the stochastic trajectory states of the emitters. For this, we focus on the \emph{averaged trajectory entanglement entropy} of a subsystem of size $l<N$, after having registered $0 \leq k \leq M$ clicks in the output arms of a network $U$, evaluated as
\begin{equation}
\label{eq:S_av}
\overline{S}^{(U)}_M(l,k) = \frac{1}{N_s}\sum_{i=1}^{N_s} S(l)\Big[\ket{ \psi^{(U)}_M(k)}_i\Big],
\end{equation}
with $N_s$ the number of samples taken and $\ket{\psi^{(U)}_M(k)}_i \nobreak\propto \nobreak c_{m_k}\dots c_{m_1}\ket{\psi_0(M,N)}$, i.e.~the state after some sequence $\vec{m}$ of $k$ detected jumps $c_{m_j}$ \eqref{eq:jump_U}. Furthermore, $S(l)[\ket{ \psi }] = - \text{Tr} \big[ \rho_\mathcal{A} \log \rho_\mathcal{A} \big]$ is the von Neumann entanglement entropy of state $\ket{\psi }$, with $\rho_\mathcal{A} = \text{Tr}_{\mathcal{B}} \ketbra{\psi}$ the reduced density matrix of subsystem $\mathcal{A}$, containing $l$ adjacent sites starting from the boundary, and $\mathcal{B}$ containing the remaining $N-l$ sites.

From a photonic perspective, an equivalent state $\ket{\psi^{(U)}_M(k)}_i$ can be obtained by subtracting $k$ single photons from the $M$-photon wavefunction at the output ports $(m_1,\dots m_k)$ from $U$ and sending the remaining $M-k$ photons back through $U$.

By sampling stochastic trajectories using matrix-product states (MPS) \cite{perez2007matrix}, we show in Fig.~\ref{fig:Fig1}(b) that when $U$ is drawn from the Haar measure, a volume-law scaling for entanglement entropy is observed, as seen in the inset. In this case, each new jump $c_i$ \eqref{eq:jump_U} generally has a nonzero overlap with any $\sigma_j^-$ and will induce long-range entanglement between all emitters in the chain. Yet, the initial growth of entanglement is upper bounded by $\overline{S}^{(U)}_M(N/2,k=1) \leq \log 2$, independent of $N$, which is obtained from the concavity of entanglement entropy \cite{divincenzo1998entanglement} (see Ref.\,\cite{supp} for details).

\textit{LON and the sampling procedure.}---In what follows, we restrict to the case $N=M$, i.e. all $M$ emitters are initialized in the excited state $\ket{\psi_N(k=0)} = \ket{\uparrow \uparrow\dots}$. The $N\times N$ unitary $U(N,D)$ that encodes the quantum jumps is implemented through a LON that consists of $D$ staggered layers of Haar random $2\times2$ unitaries, each of which can be written as Eq.\,\eqref{eq:U2by2} [see Fig.\,\ref{fig:Fig2}(a)]. For a sufficiently deep LON, one can show that sampling instances from the LON converge to drawing the $N\times N$ unitaries from the Haar measure \cite{emerson2005convergence}.

Each instance in the sample set is obtained by (i) sampling a $U(N,D)$ and (ii) sampling a quantum trajectory, thus yielding a jump sequence $m_k$ and the corresponding stochastic series of (pure) states $\ket{\psi_N(k)}$, with $0\leq k\leq N$ the number of registered jump clicks. After repeating this procedure $N_s$ times, we obtain a set of sampled trajectories, and the averaged entanglement entropies $\overline{S}^{(D)}_N(l,k)$ for subsystem size $l$ can be evaluated, yielding the entanglement of the trajectories averaged over unitaries $U(N,D)$. 

Previously, a number of works have investigated the entanglement entropy of the $M$-photon wavefunction for Fock-state boson sampling in an $N$-mode LON. In the Haar regime, the photonic wavefunction shows  volume-law scaling of entanglement entropy when exiting the LON \cite{huang2019simulating,oh2021classical}. In this chain of two-level emitters, on the other hand, the spontaneously emitted photons themselves are short-lived (stemming from the Born-Markov approximation of the quantum trajectory approach) and we study the buildup and decay of entanglement entropy between the emitters induced by registering and applying the jumps $c_j$~\eqref{eq:jump_U}. Additionally, this also marks a significant difference with the measurement-induced phase transition studied in circuit models~\cite{nahum2017quantum,skinner2019measurement} since no projective measurements are preformed on the emitters.

\textit{Numerical results and scaling of complexity.}---The stochastic simulations were run with MPS \cite{perez2007matrix,daley2014quantum}, using the C++ package \texttt{ITensor} \cite{itensor}.

In Fig.\,\ref{fig:Fig2}(b)-(c), we first study the scaling of entanglement entropy by monitoring outputs of a LON with fixed depth $D$. The largest achieved averaged entanglement entropy $\overline{S}^{(D)}_{N,\text{max}} \equiv \text{max}_{k,l}\big[ \overline{S}^{(D)}_{N}(k,l) \big]$ shows an area-law behavior. In Fig.~\ref{fig:Fig2}(b), it is seen that $\overline{S}^{(D)}_{N,\text{max}}$ does not scale with system size for fixed $D$. This is further confirmed in Fig.~\ref{fig:Fig2}(c) for subsystem scaling for the case $N=100$, where it is seen that $\overline{S}^{(D)}_{N}(k,l)$ converges to a finite value in the bulk. Note that, for any $k$, the maximal $\overline{S}^{(D)}_{N}(k,l)$ is always reached for $l=N/2$. 

Intuitively, after detecting a click from a jump $c_j$ when $D=\text{const}$, an observer can pinpoint a subset of adjacent emitters of size $2D$ where the decay could have originated from, independent of $N$. Therefore, registering a click can only generate local entanglement in the chain. LONs of fixed depth $D\ll N$ are represented by a unitary $U$ that is formulated as a banded matrix of width $2D$. Interestingly, there exist polynomial-time algorithms to efficiently evaluate $\text{Per}\big(U_T\big)$ of banded matrices, which encode output probabilities of outcomes with few or no collisions via Eq.\,\eqref{eq:boson_sampling} ~\cite{temme2012efficient,Cifuentes2016,lundow2022efficient}. The efficient evaluation of the output probabilities is in line with our result: the area law of entanglement entropy ensures that the output configurations $P(\vec{n})$ can be efficiently sampled using MPS of fixed maximal bond dimension to represent the quantum state of the emitters after $k$ clicks \cite{perez2007matrix}. 

As shown in Fig.~\ref{fig:Fig2}(d)-(e), the situation drastically changes when the network depth $D$ scales linearly with system size: $D=pN$. In Fig.~\ref{fig:Fig2}(d), we show the maximal averaged entanglement entropy $\overline{S}^{(pN)}_{N,\text{max}}$, which now has a clear linear dependence on system size $N$, thus establishing a \emph{volume law}. The simulation quickly gets out of reach for efficient simulation with MPS of a given maximal bond dimension $\chi_\text{max}$ (set to $\chi_\text{max}=700$). Also, the entanglement profiles of subsystem size $l$, shown in Fig.~\ref{fig:Fig2}(e), acquire a strong dependence on subsystem size $l$ when $p$ is increased, which we identify as volume-law for the scaling for subsystem entanglement entropy. As $p$ increases, the entanglement entropy approaches the value obtained by sampling $U$ from the $N\times N$ Haar measure (black dashed line in Fig.~\ref{fig:Fig2}(d)-(e)). 

In order to secure the classical sampling hardness, the original proof for Fock-state boson sampling requires that $N = \mathcal{O}(M^2)$ to ensure \emph{collision-free} samples \cite{aaronson2011computational}. While we are not in that regime, to our knowledge no efficient classical algorithm is known to sample the jump outcomes if $N=M$ and $D \propto N$. In our unraveling picture, we face a correlation in complexity: the entanglement entropy between the emitters in the trajectory states has volume-law scaling and quickly surpasses the limit of efficient simulation with MPS. 

On the contrary, when the trajectory-averaged entanglement entropy scales as an area law, the sample complexity (the number of trajectory states required in order to accurately sample the density matrix) may be expected to increase exponentially.
This is captured by the scaling of the (classical) Shannon entropy of the distribution over quantum trajectory states. Hence, there is a trade-off between sample complexity of trajectories and the complexity of simulating each trajectory.
It might be possible to practically exploit this trade-off in a classical algorithm, see Ref. \cite{supp} for a more detailed explanation.

\textit{Conclusions and outlook.}---It was illustrated that changing the unraveling of a straightforward, uncoupled master equation of emitters may cause drastic changes in both the entanglement of stochastic trajectory states and the sampling hardness of jump outcomes. Moreover, changing the unraveling is immediately related to an observer monitoring the decay clicks in the output arms of a LON, resulting in the unitary mixing of the decay jumps. Sampling the jump outcomes in the established monitoring scheme is equivalent to the problem of Fock-state boson sampling. Finally, a connection was established between the scaling of entanglement entropy between emitters and the classical hardness of sampling the jump outcomes.

While we have reported different scaling behavior for the trajectory entanglement entropy, we have not yet seen a conclusive signature of a scaling transition for the trajectory entanglement entropy across a critical point, such as presented in, e.g.~\cite{alberton2021entanglement,van2021entanglement}. For example, one can investigate fermionic or Gaussian models to access larger systems for the scaling analysis. 

\textit{Note added.}---While finalizing these results, we became aware of a recent work, where an entanglement scaling transition was reported in a homodyne monitoring scheme \cite{vovk2021entanglement}.

\begin{acknowledgments}
\textit{Acknowledgments}.---We acknowledge stimulating discussions with Alireza Seif and Dominik Hangleiter. M.V.R., H.D., and M.H. were sponsored by
ARO W911NF2010232, AFOSR FA9550-19-1-0399, NSF OMA-2120757, QSA-DOE and Simons foundation. L.P.G.P.~and A.V.G.~acknowledge funding by the DoE ASCR Accelerated Research in Quantum Computing program (award No.~DE-SC0020312), DARPA SAVaNT ADVENT, NSF QLCI (award No.~OMA-2120757), DoE QSA, ARO MURI, DoE ASCR Quantum Testbed Pathfinder program (award No.~DE-SC0019040), NSF PFCQC program, AFOSR, AFOSR MURI, and U.S.~Department of Energy Award No.~DE-SC0019449.  A.D.~acknowledges support from the National Science Foundation RAISE-TAQS 1839204. The Institute for Quantum Information and Matter is an NSF Physics Frontiers Center PHY-1733907.
This work used the Extreme Science and Engineering Discovery Environment (XSEDE), supported by National Science Foundation grant number ACI-1548562 and ACI-1928147, at the Pittsburgh Supercomputing Center (PSC) through allocation number TG-DMR200037 \cite{xsede}.
\end{acknowledgments}

\bibliography{refs}

\newpage 
\clearpage
\widetext
\appendix

\beginsupplement
\section{\Large Supplementary material}
\subsection{S1: The unitary mixing of jump operators: master equation vs. quantum trajectories}
In this section, we illustrate in more detail the invariance of a master equation for linearly mixing jumps with a unitary, Eq. (2) from main text, and how this invariance breaks down in the trajectory picture.

The master equation of a dissipative system is given by ($\hbar=1$)
\begin{equation}
\label{eq-app:mastereq}
    \partial_t \rho = -i[H,\rho] + \sum_j \Big( c_j \rho c^\dagger_j -\frac 12 \{ c^\dagger_j c_j, \rho \} \Big),
\end{equation}
and the trajectories for a given unraveling obey the stochastic equation~\cite{Jacobs_2006,wiseman2009quantum}
\begin{align}
\label{eq-app:unraveling}
    d \rho_\xi = -i[H,\rho_\xi] dt + \sum_j \Big( \langle c_j^\dagger c_j \rangle \rho_\xi -  \frac{1}{2} \{ c^\dagger_j c_j, \rho_\xi \}    \Big) dt  + \sum_j \left(  \frac{ c_j \rho_\xi c_j^\dag}{\langle c_j^\dagger c_j \rangle} - \rho_\xi   \right)  dN_j.
\end{align}
Here, $\xi$ labels a given realization of the stochastic variables $dN_j$, and $\rho_\xi = \ket{ \psi } \! \langle \psi |$ is the stochastic pure state of the system in each realization. In the jump picture, it holds that $dN_j = 1$ with probability $\langle c_j^\dagger c_j \rangle dt$ and zero otherwise, 
so that $dN_j^2 = dN_j$~\cite{Jacobs_2006,wiseman2009quantum}.
Using that $\rho = \overline{\rho_\xi}$ and that $\overline{dN_j} = \langle c_j^\dagger c_j \rangle dt$, where $\overline{f}$ denotes the average of $f$ over the stochastic noise terms, one immediately recovers the Lindlbdlad master equation.

A direct calculation shows that all terms in the Lindblad master equation are invariant under $c_j \rightarrow c_j' = \sum_k U_{jk} c_k$. Indeed, using $\sum_j U_{lj}^* U_{kj} = \delta_{lk}$, we find
\begin{align}
    \sum_j c_j' \rho c_j'^{\dag} =  \sum_j \sum_{kl} U_{jk} c_k \rho U_{jl}^* c_l^\dag = \sum_{kl} \delta_{lk} c_k \rho c_l^\dag = \sum_k c_k \rho c_k^\dag,
\end{align}
and similarly for the other term $\propto \sum_j\{c_j^\dagger c_j,\rho\}$. 

Importantly, this invariance does not hold generally for Eq.~\eqref{eq-app:unraveling} due to the last term, and this is precisely what motivates us to consider nonlinear trajectory-state quantities, such as the entanglement entropy. On the other hand, the probability of observing a quantum jump in time interval $dt$, $p_\text{jump}(t)= \sum_i \langle c^\dagger_i c_i\rangle dt$, is left invariant under the unitary transformation, which is why we discarded the explicit time dimension in the text and used the number of registered clicks instead. Recovering the explicit time dimension from an obtained jump-click trajectory is straightforwardly achieved by sampling the waiting times between clicks from the corresponding Poisson distributions.

\subsection{S2: Probability of observing a jump sequence}
To verify the equivalence with boson sampling, we need to evaluate the probability of observing a Markovian sequence of clicks $(m_1\dots m_M)$, Eq. (4) in main text. Alternatively, the probability of such a sequence can be obtained by evaluating the temporal correlation function of the photonic state at the output ports of the LON $U$---see Ref.~\cite{tamma2016multi}. In this section, we derive in more detail how it naturally comes out of the quantum trajectory picture of the emitters.

Starting from the state $\big|\psi_0(M,N)\big\rangle = \big|\uparrow_M \downarrow_{N-M}\big\rangle$, when $k\leq M$ jumps have been detected in the sequence $(m_1,\dots,m_k)$, we know that the quantum state of the emitters is given by
\begin{equation}
\label{eq:k_state}
    \big|\psi \big\rangle_{m_1,\dots,m_k} = \mathcal{N}(m_1,\dots,m_k)\, c_{m_k}\dots c_{m_1}\big|\psi_0(M,N)\big\rangle,
\end{equation}
with the norm
\begin{equation}
    \big|\mathcal{N}(m_1,\dots,m_k)\big|^2 = 1/\langle \psi_0(M,N)\big| c_{m_1}^\dagger\dots c_{m_k}^\dagger c_{m_k}\dots c_{m_1}\big|\psi_0(M,N)\big\rangle.
\end{equation}
The probability of sampling $c_{k+1}$ as the next jump, conditioned upon having observed the previous sequence $(m_1,\dots,m_k)$, is then obtained as the conditional probability
\begin{eqnarray}
\label{eq:k_prob}
\nonumber
    P\big(m_{k+1}|m_1,\dots,m_k\big) &=& \frac{\big \langle \psi \big| c_{k+1}^\dagger c_{k+1} \big|\psi \big \rangle_{m_1,\dots,m_k}}{\sum_i \big \langle \psi \big| c_i^\dagger c_i \big|\psi \big \rangle_{m_1,\dots,m_k}} \\
    &=&\frac{\big|\mathcal{N}(m_1,\dots,m_k)\big|^2}{M-k}\,\big \langle \psi(M,N) \big| c^\dagger_{m_1}\dots c^\dagger_{m_k} \big(c_{k+1}^\dagger c_{k+1}\big) c_{m_k}\dots c_{m_1} \big|\psi(M,N) \big \rangle,
\end{eqnarray}
where the last step follows from $\sum_i \big \langle \psi \big| c_i^\dagger c_i \big|\psi \big \rangle_{m_1,\dots,m_k} = \sum_i \big \langle \psi \big| \sigma_i^+ \sigma_i^- \big|\psi \big \rangle_{m_1,\dots,m_k}$ with the unitary transformation given by Eq.~\eqref{eq:jump_U} in main text. From this, we can evaluate the probability of observing a sequence as a product of conditional probabilities
\begin{equation}
    P(m_1,\dots m_M) = P(m_1)\times P\big(m_2|m_1\big)\times \dots \times P\big(m_M|m_1,\dots, m_{M-1}\big).
\end{equation}
Using Eq. \eqref{eq:k_prob} for the different conditional probabilities, we arrive at Eq.~\eqref{eq:markovian_clicks} from the main text.

\subsection{S3: Bound on initial entanglement growth}

In this section, we formulate a universal upper bound for the initial growth rate of entanglement entropy when $M=N$, i.e. when all emitters start in the excited state, and $U_{N\times N}$ is drawn from the Haar measure. 

For this, we know that, after registering one jump $c_i=\sum_j U_{ij} \sigma^-_j$, the quantum state of emitters is given by
\begin{equation}
\label{eq:psi1}
    |\psi_1(N) \rangle = c_i |\psi_0(N)\rangle = U_{i1}|\downarrow \uparrow \uparrow \uparrow\dots \rangle + U_{i2}| \uparrow \downarrow \uparrow \uparrow\dots \rangle + U_{i3}|\uparrow \uparrow \downarrow \uparrow \dots \rangle + \dots.
\end{equation}
For this state, we can compute the reduced density matrix of a subsystem $\mathcal{A}$, composed of $l$ sites, by tracing out the environment $\mathcal{B}$ composed of $N-l$ sites, 
\begin{equation}
    \rho_\mathcal{A} = \text{tr}_\mathcal{B} \big[ |\psi_1(N) \rangle \langle\psi_1(N) | \big]
    = p |\psi_{\mathcal{A},0}\rangle \langle \psi_{\mathcal{A},0}| + (1-p) |\psi_{\mathcal{A},1}\rangle \langle \psi_{\mathcal{A},1}|.
    \label{eq:rhoA}
\end{equation}
Here, $p=\sum_{j\in \mathcal{A}} \big|U_{ij}\big|^2$ is the probability of finding the de-excitation in subsystem $\mathcal{A}$. Furthermore, $|\psi_{\mathcal{A},0}\rangle  = |\uparrow_{l\in\mathcal{A}}\rangle$ is the quantum state when the jump is detected in the environment, and $|\psi_{\mathcal{A},1}\rangle   \nobreak = \nobreak 1/\sqrt{p}\sum_{j\in\mathcal{A}} U_{ij} |\downarrow_j \uparrow_{l\neq j}\rangle$ the state if the jump occurs in $\mathcal{A}$. Hence, the reduced density matrix $\rho_\mathcal{A}$ is composed of a statistical mixture of two pure and orthogonal quantum states with a classical probability $p$, for which the entanglement entropy equals $S(p) = -p\log p - (1-p)\log{(1-p)}$.

Since the Von Neumann entanglement entropy is a concave function, meaning that $S\big(\sum_j \lambda_j \rho_j \big) \nobreak \geq \nobreak \sum_j \lambda_j S\big(\rho_j\big)$ for some statistical ensemble of density matrices $\rho_j$ with probabilities $\lambda_j$ (see e.g. Ref. \cite{divincenzo1998entanglement}), an upper bound can be found by evaluating the entanglement entropy of the averaged density matrix of an ensemble. If we average instances from the $N\times N$ Haar measure in \eqref{eq:rhoA}, we know that $\mathbb E_U \big[|U_{ij}|^2\big]=1/N$ and therefore that $\mathbb E_U[p]=l/N$, with $\mathbb E_U[\cdot]$ denoting the average over the measure of $N\times N$ Haar unitaries. Using the averaged probability for the statistical mixture given in \eqref{eq:rhoA} after registering one click, a bound is found on the averaged entanglement entropy of a subsystem containing $l\leq N$ sites,
\begin{equation}
\label{eq:bound_profile}
\overline{S}_N(l, k=1) \leq -\frac{l}{N}\log{\frac{l}{N}} - \frac{N-l}{N} \log{ \frac{N-l}{N} }.
\end{equation}
This means that we find a universal bound for $x=l/N$ as $\overline{S}(x,k=1)\leq h(x) \equiv -x\log x - (1-x)\log{(1-x)}$. Numerically, we find that the bound also holds later, so that $\overline{S}(x,k) \leq k h(x)$.

\begin{figure}
    \centering
    \includegraphics{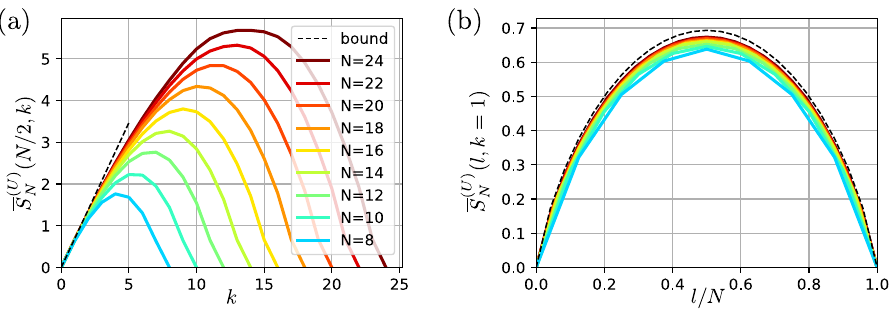}
    \caption{A comparison between the entanglement generated when $U$ is sampled from the $N\times N$ Haar measure; the analytical upper bound \eqref{eq:bound_profile} is almost saturated. (a) Evolution of half-chain (maximal) entanglement entropy after detection of $k$ clicks, for various $N$. (b) Entanglement subsystem profile after registering one jump click, same color codes as panel (a).}
    \label{fig:bound}
\end{figure}

In Fig.~\ref{fig:bound}, we illustrate how bound \eqref{eq:bound_profile} is approached. The initial growth for $\overline{S}(N/2,k)$ lies close to the bound, as shown in Fig.~\ref{fig:bound}(a). In Fig.~\ref{fig:bound}(b), we illustrate that the bound from \eqref{eq:bound_profile} for the bipartite entanglement $\overline{S}(x,k=1)$ is approached when $N$ is increased. 

Note also that the top of the curve $\overline{S}_\text{max}$ seems to be slightly flattened for $N=24$. The maximal bond dimension, set to $\chi_\text{max}=700$ for the MPS simulation, was not sufficient to capture all statistical fluctuations of entanglement entropy. Therefore the data for $N=24$ was left out for $S_\text{max}$ in Fig. 1b in the main text. We checked different sample trajectories to ensure that all data points $N\leq22$ were not suffering from this issue.

\subsection{S4: Scaling of the average entropy and statistics of the unraveling}
We mention in the main text, just before conclusions and outlook, that inducing entanglement in the stochastic trajectories reduces the statistical trajectory fluctuations for sampling the averaged density matrix. In this section, we explain this in more detail for the case of $N$ emitters from the main text.

Let $\overline{\rho}_l = \sum_j \lambda_j \trace_{N-l}\left[\ket{\psi^j}\!\bra{\psi^j} \right]$ denote the trajectory-averaged state of a subsystem of $l$ sites, where $\lambda_j$ is the probability with which state $\ket{\psi^j}$ occurs in the ensemble of trajectories, with $\sum_j \lambda_j = 1$. The trajectory-averaged von Neumann entropy of the subsystem, $\overline{S} = \sum_j \lambda_j S\left( \trace_{N-l}\left[\ket{\psi^j}\!\bra{\psi^j} \right] \right)$, satisfies
\begin{align}
\label{eq:entropybounds}
  \overline{S} \leq  S(\overline{\rho}_l) \leq \overline{S} + \mathcal{H}(\lambda),
\end{align}
where $\mathcal{H}(\lambda) = - \sum_j \lambda_j \ln \lambda_j$ is the (classical) Shannon entropy of the distribution $\{ \lambda_j \}$ that characterizes the mixture~\cite{dalibard1992wave,NielsenChuang}.

In the main text, we numerically studied $\overline{S}$. The entropy of the trajectory-averaged state is easy to obtain from the fact that, at the level of the master equation, each emitter remains in the excited state with a probability $p(t) = e^{-\gamma t}$. The entropy of $l$ emitters is then
\begin{align}
    S(\overline{\rho}_l;t) = l \times \Big( p(t) \log{p(t)} + (1-p(t)) \log{(1- p(t))} \Big).
\end{align}
This entropy satisfies a volume law, $S(\overline{\rho}_l) \propto l$.

Using Eq.~\eqref{eq:entropybounds}, we find that, for the classical Shannon entropy of the mixture, 
\begin{align}
\label{eq:bound}
    \mathcal{H}(\lambda) \geq S(\overline{\rho}_l) - \overline{S} \sim \mathcal{O}(l) - \overline{S}.
\end{align}
This implies that whenever we find that $\overline{S}$ follows an area law, or scales slower than $S(\overline{\rho}_l)$ with $l$, the classical entropy $\mathcal{H}(\lambda)$ characterizing the unravelling, must compensate for this and scale with the volume of the system, $\mathcal{H}(\lambda) \sim \mathcal{O}(l)$. For numerical purposes, the number of distinct area-law trajectories in an unraveling needed to sample a master equation leading to a volume-law density matrix should scale exponentially to satisfy bound \eqref{eq:bound}. On the other hand, using volume-law trajectories, one might reach sufficient statistical accuracy after obtaining a set of samples with polynomial (or even constant) size. We plan to investigate this issue further in a follow-up work, with the goal of discovering optimal unravelings that have a balance between quantum entanglement (hardness of classically computing a given trajectory) and the number of samples needed (hardness of classical sampling) to acquire sufficient statistical accuracy.

 \end{document}